\documentclass[
twocolumn,
aps,nofootinbib,showpacs,showkeys,preprint
tightenlines,preprintnumbers,
,superscriptaddress
] {revtex4}

\usepackage{epsf,epsfig,subfigure,graphicx,amsmath,amssymb}
\usepackage{color}
\newcommand{\dis}[1]{\begin{equation}\begin{split}#1\end{split}\end{equation}}

\def\lsim{\lower.7ex\hbox{$\;\stackrel{\textstyle<}{\sim}\;$}}

 \def\Z{{\bf Z}}

 \def\Z{{\bf Z}}
 
 \def\Zpino{{$\tilde{Z}'$}}

\def\SMg{{SU(3)$_c\times$SU(2)$_W\times$U(1)$_Y$}}
\def\LSTMM{{MM}}
\def\LSTZPM{{$\tilde{t}_l Z'$M}}

\def\MGUT{$M_{\rm GUT}$}
\def\ie{{\it i.e.}\ }

\def\two{{\bf 2}}
\def\fiveb{\overline{\bf 5}\,}
\def\five{{\bf 5}}

\def\ten{{\bf 10}}
\def\one{{\bf 1}}

\def\threeb{\overline{\bf 3}\,}
\def\three{{\bf 3}}

\begin{document}
%\draft
\preprint{SNUTP 11-008, TU-886}
\title{Gauge mediation to effective SUSY through U(1)s with a dynamical SUSY breaking, and
string compactification}
\address{Department of Physics, Tohoku University, Sendai 980-8578, Japan}
\author{Kwang Sik Jeong$^1$, Jihn E.  Kim$^{2,3}$, Min-Seok Seo}
\address{Department of Physics and Astronomy and Center for Theoretical
 Physics, Seoul National University, Seoul 151-747, Korea\\
$^{\it\tiny 3}$GIST College, Gwangju Institute of Science and Technology, Gwangju 500-712, Korea}

\begin{abstract}
We investigate the possibility of U(1)$'$ mediation, leading to an effective SUSY where the first two family sfermions are above 100 TeV but the third family sfermions and the Higgs doublets are in the TeV region(or the light stop($\tilde t_l$) case). The U(1)$'$ gaugino, {\it Zprimino} ($Z'$-ino), needs not to be at a TeV scale, but needs to be somewhat lighter than the messenger scale. We consider two cases, one the mediation is only through U(1)$'$ and the other through U(1)$'$ and the electroweak hypercharge U(1)$_Y$. In the SUSY field theory framework, we calculate the superpartner mass spectra for these two cases. We also point out that the particle species needed for these mechanisms are already obtained from a \Z$_{12-I}$ orbifold compactification.
\keywords{Effective SUSY, Light stop, U(1)$'$ mediation, Orbifold compactification}
\end{abstract}

 \pacs{12.60.Jv, 14.80.Ly, 11.25.Wx, 11.25.Mj}

 \maketitle

%%%%%%%%%%%%%%%%%%%%%%%%%%%%%%%%%%%%%%%%%%%%%%%%%%%%%%%%%%%%%%%%%%%
%%%%%%%%%%%%%%%%%%%%%%%%%%%%%%%%%%%%%%%%%%%%%%%%%%%%%%%%%%%%%%%%%%%
\section{Introduction}

Supersymmetry(SUSY) and its breaking mechanism have been the most active particle theory
research in the last three decades. In particular, the SUSY flavor problem has led to
the gauge mediated SUSY breaking (GMSB) \cite{DineNeslson,GMSBrev}.
The attractive gravity mediation scenario for transmitting SUSY breaking down to the visible sector probably violates the flavor independence of interactions, but there are ways in the gravity mediation also to suppress the flavor changing neutral couplings(FCNC) of the standard model(SM) fermions in the effective SUSY(effSUSY) framework \cite{effSUSY96}. In the effSUSY, the first two family sfermions are sufficiently heavy above 5--20 TeV while the third family sfermion masses are in the 100 GeV--1 TeV region. The SUSY flavor solution by the GMSB relies on the family independence of the sfermion interaction, for which the gauge interactions do not distinguish family members. The family independence of sfermion masses needs the dominant SUSY breaking source with color SU(3)$_c$ charge, the weak SU(2)$_W$ charge and the weak hypercharge $Y$. There exists the {\it SUSY breaking source at some hidden sector scale $\Lambda_h$ below $10^{12}$ GeV for the GMSB} is useful \cite{Kim07} and the messengers, carrying the visible sector gauge charges, acquire SUSY braking $F$ (or $D$) terms. The visible sector sfermions obtain masses via these messenger $F$ terms and sometimes the grand unification(GUT) messenger multiplets have been considered for this transmitting purpose \cite{GMSBrev}.

Even though the original GMSB seems to be attractive, similar related ideas in terms of U(1)s have been suggested by Langacker, Pas, Wang and Yavin \cite{Yavin07} and Mohapatra and Nandi, and Kikuchi and Kubo \cite{MohNandi97,Kikuchi08}. The Langacker {\it et al.} mechanism employs an extra $Z'$ gauge interaction instead of the whole \SMg\ interactions of the SUSY breaking source. The messengers and the SM fields carry the $Z'$ charges and the $\tilde{Z}'$ gaugino mass is triggering the superpartner  masses of the SM fields through the messengers. In addition, they assume a TeV scale $Z'$, but the low energy scale of which is not needed in general just for a mediation mechanism alone. On the other hand, the Mohapatra-Nandi mechanism uses U(1)$_{Y_1}$and U(1)$_{B-L}$ and both of these U(1)s participate in the breaking of SUSY and
U(1)$_{Y_1}$ to obtain the U(1)$_{Y}$ of the SM and also the transfer of SUSY breaking to the superpartners of the SM. The SUSY breaking source can be of dynamical origin as suggested by the well-known dynamical SUSY breaking(DSB) models in SO(10)$'$ with $\bf 16'$ or $\bf 16'+10'$ \cite{SO10}, or in an SU(5)$'$ model with $\bf 10'+\overline{5}\,'$ \cite{SU5}. We understand that the effective Polonyi form for SUSY breaking \cite{Polonyi} is parametrizing the DSB models. Therefore, for a full description of GMSB or {\it mixed mediation}, we should rely on the string origin of SUSY breaking endowing one hidden family of SU(5)$'$ or SO(10)$'$. There already exist models from the superstring orbifold compactification implementing the SUSY breaking source SU(5)$'$ with the visible sector \SMg\  \cite{GMSBstSU3} or with the flipped SU(5) \cite{KimKyaefiveprime09}. In particular, the one hidden sector family models of SO(10)$'$ and SU(5)$'$ cannot carry SU(3)$_c$ color and SU(2)$_W$ charges, or the hidden sector does not satisfy the one family condition. Then, the gauge mediation is better to be through U(1)s, and it is not expected that the SM families carry the same U(1)$'$ charges which does not satisfy the chief merit of the GMSB the family independence of the mediation. The best we can anticipate for low energy SUSY is an effSUSY \cite{effSUSY96} that the superpartners of two light family members are much heavier than the TeV scale.

The recent Large Hadron Collider(LHC) reports exclude squarks in the TeV region \cite{LHC11Cofs} even though these analyses are based on the $R$-parity conserving constrained MSSM. Therefore, in the SUSY framework the effSUSY is the next serious candidate to be analyzed thoroughly \cite{effPhenJeong11}.
If the $R$-parity conserved, the axino \cite{AxinoCDM} or gravitino LSP \cite{GravtinoCDM} models are not free from the LHC problem. For example, for $F_a=10^{11}$ GeV and 1 TeV squark mass, the squark decay line to the axino vertex is estimated as a few mm order, which is swamped by decays to NLSPs.

In this regard, we note that a simpler model building exists in SUSY field theory framework via the Intrilligator, Seiberg and Shih (ISS) mechanism where the vacuum is unstable but have a sufficiently long lifetime \cite{ISS,Murayama06,ISSfollow}. As noted from the string compactification, the total number of 4 dimensional(4D) chiral fields are somewhere between 100 and 200 and it is very difficult for the SUSY breaking source to carry all the \SMg\ charges. The ISS mechanism is not free of this problem, if not impossible, since for example SU($N_h$) with $N_f$ flavors needs $N_h+1\le N_f<\frac32 N_h$ chiral fields for an unstable minimum. The simplest case SU(5)$'$ needs 6 or 7 vectorlike flavors, which has been realized in string compactifications \cite{Kim07} where however the SU(2)$_W$ is broken at the hidden sector scale and the messengers do not carry the color charges. So, it is likely that the original idea of GMSB in the ISS form needing a baroque representation may not be realizable from string compactification.

On the other hand, the Langacker {\it et al.} type or the Mohapatra-Nandi type mediation, employing only U(1)s for mediation, can be easily realizable in SUSY breaking models of one family SU(5)$'$ or of ISS.

We note that there result some phenomenologically acceptable string vacua, where {\it light stop $Z'$ mediation} (LSTZPM or \LSTZPM) and  {\it light stop mixed mediation} (LSTMM or simply \LSTMM)\footnote{We pick up the light stop among the third family members because the RG evolution is dominated by the top Yukawa coupling.} to an effSUSY, from 10D string to a 4D minimal supersymmetric SM(MSSM) \cite{KimKyaegut,KimKyaeSM,ChoiKimBk,Katsuki}:
\begin{itemize}
\item {\bf Model }\LSTZPM

\item[1.] Many U(1)s may contribute in the mediation. Here we choose the simplest possibility that only one U(1)$'$ with the superpartner Zprimino ($Z'$-ino), \Zpino, is effective in the mediation.

\item[2.] The SUSY breaking source at $\Lambda_h$ does not carry the weak hypercharge $Y$, or the low energy SM does not result. The messenger sector at $M_{\rm mess}$ carries the $Z'$ charge $Y'$ but does not carry the weak hypercharge $Y$.

\item[3.] The superpartners of the third family fermions, ($t,b,\tau,\nu_\tau$) do not carry the $Z'$ charge $Y'$. This item realizes the effSUSY.

\item[4.] The Higgs doublets do not carry  the $Z'$ charge $Y'$. The SU(2)$_W\times$U(1)$_Y$ breaking is naturally achieved by a running of Higgs boson masses.

\end{itemize}

\begin{itemize}
\item {\bf Model }\LSTMM

\item[1.] Many U(1)s may contribute in the mediation. Here we choose the simplest possibility that only one U(1)$'$ is effective in addition to U(1)$_Y$ of the SM. These gauge bosons are  $Z'$ and $B$, and their superpartners are called Zprimino \Zpino\ and Bino.

\item[2.] The SUSY breaking source does not carry the weak hypercharge $Y$, or the low energy SM does not result. The messenger sector carries both the weak hypercharge $Y$ and the $Z'$ charge $Y'$.

\item[3.]  The superpartners of  the third family fermions do not carry the $Z'$ charge $Y'$.  This item realizes the effective SUSY \cite{effSUSY96}.

\item[4.] Higgs doublets do not carry  the $Z'$ charge $Y'$.

\item[5.] The SU(2)$_W\times$U(1)$_Y$ breaking is done by a fine-tuning between parameters of the Higgs boson mass matrix \cite{Yavin07}.

\end{itemize}

These two cases are the effSUSY generalization of the U(1)$'$ mediation \cite{Yavin07} and the mixed U(1)s mediation \cite{MohNandi97}. In fact, both of these cases are explicitly found in \Z$_{12-I}$ orbifold compactification \cite{GMSBstSU3}.

In Sec. \ref{sec:MedPhen}, we discuss the general features of \LSTZPM\ and \LSTMM\ on the spectra of superpartners of the SM, and in Sec. \ref{sec:string} we present such realizations from a published string compactification model \cite{GMSBstSU3}. Here the gauge symmetry breaking to the MSSM is achieved by the vacuum expectation values(VEVs) of some scalar fields obtained from the orbifold compactification. In Appendix A, we present the renormalization group(RG) running inputs and the relevant formulae. In Appendix B, we present two tables on charged and neutral singlets used in  Sec. \ref{sec:string}. Section \ref{sec:Conclusion} is a conclusion.

%%%%%%%%%%%%%%%%%%%%%%%%%%%%%%%%%%%%%%%%%%%%%%%%%%%%%%%%%%%%%%%%%%%%%%%%%%%%%%%%%%%%%%%%%%%%
\section{SUSY breaking mediation by U(1)$'$}\label{sec:MedPhen}

We argue that U(1)$'$ mediation of SUSY breaking is of general nature in string compactification.
A prototype example has been given in Ref. \cite{GMSBstSU3} where the SUSY breaking source is provided by the confining hidden sector SU(5)$'$ with one family $\ten'+\fiveb'$. In Ref. \cite{GMSBstSU3}, the original GMSB idea has been commented by assuming the Planck scale singlet VEVs, but it is probable that some needed singlets do not have that large VEVs. Then, the unremovable SUSY breaking mediation is through U(1)s. This explicit string model will be commented after we present phenomenological aspects of \LSTZPM\ and \LSTMM\ schemes.

\begin{figure}[t]
  \begin{center}
  \begin{tabular}{c}
   \includegraphics[width=0.4\textwidth]{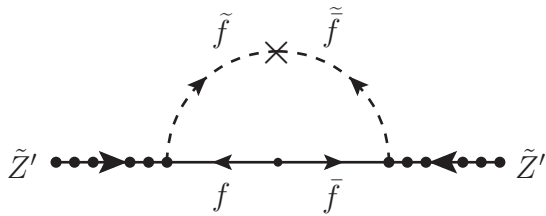}
   \end{tabular}
  \end{center}
  \caption{The mass diagram of Zprimino. The SUSY breaking insertion from DSB is $\times$. The bulleted line is \Zpino. This soft mass is added to the SUSY mass.}
\label{fig:ZpMass}
\end{figure}

We consider two U(1) gauge bosons, $B_\mu$ corresponding to $Y$ of the SM and $Z'_\mu$ corresponding to an additional hypercharge $Y'$. The messenger matter fields $f$ and $\bar f$ have the quantum numbers of $f(Y,Y')$ and $\bar f(-Y,-Y')$. Both of these possibilities are possible with the model of Ref. \cite{GMSBstSU3}. In these models, the Zprimino mass diagram appears as in Fig. \ref{fig:ZpMass}. We emphasize the $Z'$ mediation by representing $Z'$ as a bulleted one and Zprimino as a bulleted line.

%%%%%%%%%%%%%%%%%%%%%%%%%%%%%%%%%%%%%%%%%%%%%%%%%%%%%%%%%%%%%%%%%%%%%%%%%%%%%%%%%%%

%%%%%%%%%%%%%%%%%%%%%%%%%%%%%%%%%%%%%%%%%%%%
\subsection{Light stop $Z'$ mediation}

\begin{figure}[b]
  \begin{center}
  \begin{tabular}{c}
   \includegraphics[width=0.4\textwidth]{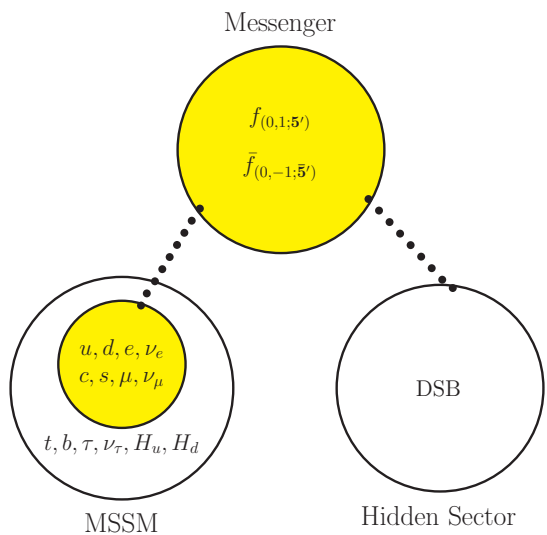}
   \end{tabular}
  \end{center}
  \caption{An effSUSY through \LSTZPM.  The $Z'$ line is the bulleted one.}
\label{fig:LSTZPM}
\end{figure}

The messenger fields, carrying the hidden sector color such as the SU(5)$'$ charge have the following $(Y,Y';\rm SU(5)')$
\dis{
f(0,1;\five'),~~ \bar f(0,-1;\fiveb')
}
and the third family members do not carry the $Y'$ charge. In addition, Higgs doublets also do not carry the $Y'$ charge. In this case, the mediation mechanism is shown pictorially in Fig. \ref{fig:LSTZPM}, which is called \LSTZPM. In this case, a light Higgs boson and the light 3rd family members are obtained naturally. In Fig. \ref{fig:LSTZPM}, the U(1)$'$ charged sectors are colored yellow.
%%%%%%%%%%%%%%%%%%%%%%%%%%%%%%%%%%%%%%%%%%%%%%%%%%%%%%%%%%%%%%%%%%%%%%%%%%%%%%%%%%%

As a field theory example, we consider an anomaly free U(1)$'$ charge assignment as $Y'=B-L$ for the first two families, and $Y'=0$ for the third family members as listed in Table \ref{table:BminusL12}. Certainly, it may be difficult for this model to produce a successful flavor structure if the U(1)$'$ breaking scale is below 10$^{12}$ GeV and the messenger scale is at the GUT scale. So, we assume the messenger scale is low, \ie only a factor of 100 larger than the DSB scale. Here, our main concern is obtaining the superparticle spectrum.
%%%%%%%%%%%%%%%%%%%%%%%%%%%%%%%%%%%%%%%%%%%%%%%%%%%%%%%%%%%%%%%%%
 \begin{table}[t]
\begin{center}
\begin{tabular}{|c|cc||c|cc|}
\hline  light families &  $Y$ &$Y'$ & 3rd family and $H_{d,u}$&  $Y$ &$Y'$
\\[0.2em]
\hline &&&&&\\ [-1.1em]
$q_{1,2}$ & $\frac16$ & $\frac{1}3$ &$(t,b)$ & $\frac16$ & 0
\\[0.4em]
$u^c_{1,2}$& $\frac{-2}3$ & $\frac{-1}3$& $t^c$& $\frac{-2}3$ & 0
\\[0.4em]
$d^c_{1,2}$& $\frac{1}3$ & $\frac{-1}3$&$b^c$& $\frac{1}3$ & 0
\\[0.4em]
$l_{1,2}$& $\frac{-1}2$ & $-1$&$(\nu_\tau,\tau)$& $\frac{-1}2$ & 0
\\[0.4em]
$e^c_{1,2}$& $1$ & $1$&$\tau^c$& $1$ & 0
\\[0.4em]
$N^c_{1,2}$& $0$ & $1$& $N^c_{3}$& $0$ & 0
\\[0.4em]
& & & $H_d$& $\frac{-1}{2}$ & 0
\\[0.4em]
& & & $H_u$& $\frac{1}{2}$ & 0
\\[0.4em]
\hline
\end{tabular}
\end{center}
\caption{The $Y'$ charges of the SM fermions, Higgs doublets and heavy neutrinos.} \label{table:BminusL12}
\end{table}

The Zprimino $\tilde{Z}'$ obtains mass through the diagram shown in Fig. \ref{fig:ZpMass},
where the SUSY breaking insertion is shown as $\times$.
Below the messenger mass scale $M_{\rm mess}$, the Zprimino soft mass is estimated as\footnote{This soft mass is added to the supersymmetric mass.}
\dis{
\frac{M_{\tilde Z^\prime}(\mu)}{g^2_{Y^\prime}(\mu)}
= -\frac{1}{8\pi^2}\frac{F_{\rm mess}}{M_{\rm mess}},
}
where $F_{\rm mess}$ is the relevant $F$-term of the messenger sector.

\begin{figure}[t]
  \begin{center}
  \begin{tabular}{c}
   \includegraphics[width=0.4\textwidth]{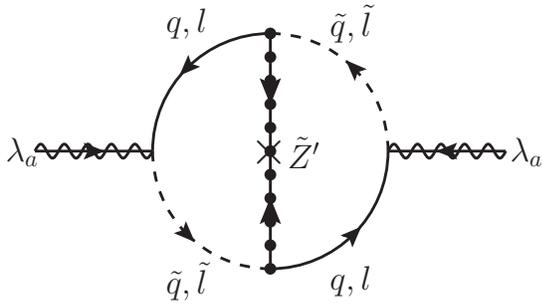}
   \end{tabular}
  \end{center}
  \caption{The mass diagram of the SM gauginos. The SUSY breaking from Zprimino sector is  shown as $\times$.  The $\tilde{Z}'$ line is a bulleted line.}
\label{fig:SMgaugino}
\end{figure}

Since the messengers are not charged under the SM gauge group, the MSSM gaugino
masses are induced only through RG running from the loops shown
in Fig. \ref{fig:SMgaugino} where the Zprimino mass is shown as $\times$.
Assuming that U(1)$^\prime$ is broken at a scale much larger
than $M_{\tilde Z^\prime}$ but below $M_{\rm mess}$, one can obtain
\dis{
\frac{M_a(\mu)}{g^2_a(\mu)} =
-\frac{c_a g^2_{Y^\prime}(M^\prime)}{(8\pi^2)^2}
M_{\tilde Z^\prime}(M^\prime) \ln\left(\frac{M_{\rm mess}}{M^\prime}\right),
}
for $\mu<M^\prime$ with $M^\prime$ being the U(1)$^\prime$ breaking scale.
For the U(1)$^\prime$ charge assignment given in Table \ref{table:BminusL12},
$c_a$ are given by
\dis{
c_Y=\frac{92}{27}, \quad c_2=\frac{8}{3}, \quad c_3=\frac{8}{9},
}
and thus the MSSM gauginos have a compressed mass spectra compared to the
ordinary gauge mediation.

\begin{figure}[h]
  \begin{center}
  \begin{tabular}{c}
   \includegraphics[width=0.4\textwidth]{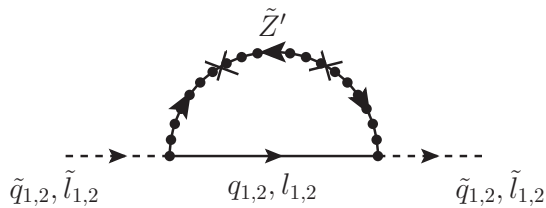}
   \end{tabular}
  \end{center}
  \caption{The first two family sfermion($\tilde{q}_{1,2},\tilde{l}_{1,2}$) mass diagrams. The SUSY breaking from Zprimino sector is  shown as $\times$. }
\label{fig:Sflight}
\end{figure}

On the other hand, the first two family sfermions directly couple to
U(1)$^\prime$ and obtain masses as
\dis{
m^2_{\tilde q_{1,2},\tilde l_{1,2}} = Y^{\prime 2}_{q_{1,2},l_{1,2}}
M^2_{\tilde Z^\prime},
}
at the messenger scale.
The dominant effect on the RG running of their masses comes from the loops
involving the Zprimino as shown in Fig. \ref{fig:Sflight}.

%%%%%%%%%%%%%%%%%%%%%%%%%%%%%%%%%%%%%%%%%%%%%%%%%%%%%%%%%%%%%%%%%%%%%%%%%%%%%%%%%%%

%%%%%%%%%%%%%%%%%%%%%%%%%%%%%%%%%%%%%%%%%%%%%%%%%%%%%%%%%%%%%%%%%%%%%%%%%%%%%%%%%%%

\begin{figure}[h]
  \begin{center}
  \begin{tabular}{c}
   \includegraphics[width=0.4\textwidth]{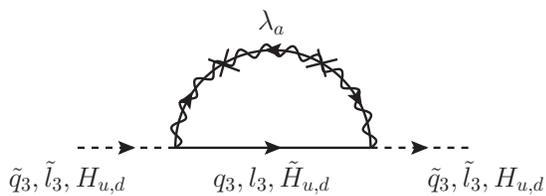}
   \end{tabular}
  \end{center}
  \caption{The mass diagrams for the third family sfermion($q_3,l_3$) and Higgs bosons. The SUSY breaking from the SM gauginos are shown as $\times$.  }
\label{fig:sqHiggs}
\end{figure}

Due to the desired gauge coupling hierarchy, the MSSM gauginos are lighter than the first two family sfermions. The third family sfermions obtain mass through the diagram shown in Fig. \ref{fig:sqHiggs}, where the SUSY breaking mass of the MSSM gauginos are shown as $\times$.
The soft scalar masses for the third family sfermions and Higgs bosons can be obtained from
the RG running equations, which are the same as those in the MSSM at the leading order.

The electroweak symmetry breaking is achieved radiatively \cite{Ibanez82} by the RG running equations.
For a successful electroweak symmetry breaking in SUSY models, we need a TeV scale $\mu$ term. Because it is a superpotential term, it is not generated radiatively. In the GMSBs and in our \LSTZPM\ and \LSTMM, we need to introduce it independently. The gravitational (or more explicitly in string models the moduli) interactions introduce a nonrenormalizable term of the form for the Higgsino doublet pair \cite{Mu},
\dis{
\sim\frac{S_1S_2}{M_P}H_{u} H_{d}\label{eq:KimNilMu}
}
where $M_P\simeq 2.44\times 10^{18}$ GeV, and $S_{1,2}$ are the SM singlet(s). With VEVs of  $S_{1,2}$ in the $10^{10-11}$ GeV region, we obtain the needed magnitude of $\mu$. Without this additional gravity effect, it may be difficult to obtain a successful electroweak symmetry breaking if not impossible. The Giudice-Masiero mechanism \cite{Giudice} does not introduce a right order of $\mu$ from the K\"ahler potential since the gravitino mass $m_{3/2}$ is required to be much smaller than the electroweak scale.

Below, we introduce the needed $\mu$ independently from the \LSTZPM\ and \LSTMM.

%%%%%%%%%%%%%%%%%%%%%%%%%%%%%%%%%%%%%%%%%%%%%%%%%%%%%%%%%%%%%%%%%%%%%%%%%%%%%%%%%
\begin{figure}[t]
\begin{center}
\begin{minipage}{15cm}\hskip -7cm
   \includegraphics[width=0.5\textwidth]{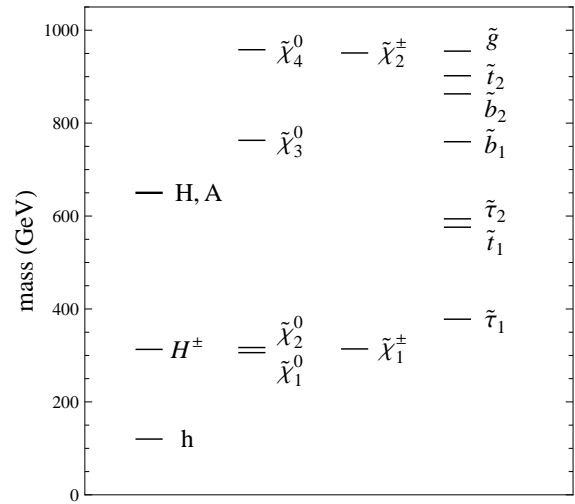}
%\centerline{
%{\hspace*{.2cm}\epsfig{figure=figSpectrum.eps,angle=0,width=10cm}}
%}
\end{minipage}
\end{center}
\caption{The sparticle and Higgs boson mass spectra in the \LSTZPM.
We have taken $M_{\rm mess}=10^{14}$ GeV, $M_{Z^\prime}=10^8$ GeV and
$M_{\tilde Z^\prime}(M_{\rm mess})=1.8\times 10^6$ GeV, for which the squark masses of the first two families are above $10^6$ GeV.}
\label{fig:SpectrumZp}
\end{figure}

We present the spectra based on Table \ref{table:BminusL12} in Fig. \ref{fig:SpectrumZp}.\footnote{But in the string example discussed in Sec. \ref{sec:string}, since the first and second generation SU(2)$_L$ doublet quarks and leptons are not charged under U(1)$'$, the spectra may be distinct from Fig. \ref{fig:SpectrumZp}.}
From Fig. \ref{fig:SpectrumZp}, we note that the lightest Higgs boson mass is around 120 GeV. We also note that the stop mass is near 580~GeV which is lower than the recent CMS bound of 1.2~TeV \cite{LHC11Cofs}. However, the latter is not a serious problem, for we can achieve this CMS bound by enlarging the Zprimino mass,\footnote{
To get $m_{\tilde t_1}\gtrsim 1.2$ TeV in the example, one can take $M_{\tilde Z^\prime}\gtrsim 3.8\times 10^6$ GeV at
the messenger scale.
The electroweak symmetry breaking would then require a rather large Higgs $\mu$ term:
$\mu \gtrsim 670$ GeV.
}
which will subsequently raise the MSSM gaugino masses.
Note that the third family sfermions and Higgs bosons acquire soft masses through RG running
from the loops involving MSSM gauginos.
This explains why they are lighter than the MSSM gauginos.

In the example, $\tilde \chi^0_4$ and $\tilde \chi^\pm_2$ come mostly from the neutral and
charged wino, respectively, while $\tilde \chi^0_3$ is bino-like.
The lightest ordinary sparticle is $\tilde \chi_1^0$, which is higgsino-like and has mass
around 306 GeV. Since the gluino mass is comparable to the wino/bino mass, the third generation squarks are not so heavy at high energy scales. As a result, $m^2_{H_u}$ is slowly driven to negative as the energy scale goes down compared to the ordinary gauge mediation,
and a small $\mu$-term is required for the electroweak symmetry breaking.
In the example, $\mu=313$ GeV and $B=133$ GeV at the weak scale, and $\tan\beta=10$.

Meanwhile, the lightest SUSY particle(LSP) is given by the gravitino having mass $\sim \Lambda^3_h/M^2_{Pl}$.

%%%%%%%%%%%%%%%%%%%%%%%%%%%%%%%%%%%%%%%%%%%%
\subsection{Mixed mediation}

The only difference of \LSTMM\ from the \LSTZPM\ is that the messenger fields carry the $Y$ charge,
\dis{
f(Y_f,1;\five'),~~ \bar f(-Y_f,-1;\fiveb').
}
If the bino is much heavier than other MSSM gauginos, the top Yukawa interaction
would drive not only $m^2_{H_u}$ but also the left-handed stop mass squared
to negative at around the weak scale.\footnote{
In the situation under consideration, the left-handed stop and up-type Higgs boson
would acquire soft masses as
\dis{
&m^2_{H_u}(\mu) \simeq \frac{1}{4} g^2_Y P_1
+ \frac{1}{2} g^2_Y P_2 - 3y^2_t P_3, \\
&m^2_{\tilde t_L}(\mu) \simeq \frac{1}{36} g^2_Y P_1
+ \frac{1}{18} g^2_Y P_2 - y^2_t P_3,
}
at $\mu<M_{Z^\prime}$.
Here $P_{1,2,3}$ are positive numbers of ${\cal O}(M^2_{\tilde B})$.
Also note that the first term corresponds to the bino-mediated contribution at $M_{\rm mess}$,
while the latter two are from the RG effects associated with $U(1)_Y$ gauge and top-Yukawa
coupling, respectively.
It is obvious that $m^2_{\tilde t_L}<m^2_{H_u}$ at a low energy scale.
}
To avoid such a problem in the \LSTMM\ scenario, we need $Y^2_f g^2_Y \lesssim 10^{-3}g^2_{Y^\prime}$
so that the bino mediation induces soft scalar masses at most of the order
of the wino/gluino mass. In this case, the effective SUSY can be obtained since the bino mediation is much weaker than the $Z^\prime$ mediation. Nonetheless, the bino mediation can still change the mass spectra of light sparticles and Higgs bosons.

For the case that the bino mediation is as important as the $Z^\prime$ mediation,
a tachyonic stop can be avoided if the third family sfermions are charged under U(1)$^\prime$.
Only the wino and gluino then remain light while all the scalars acquire quite large soft masses. Hence, we need to fine-tune the Higgs mass parameters to achieve the correct electroweak symmetry breaking.

%%%%%%%%%%%%%%%%%%%%%%%%%%%%%%%%%%%%%%%%%%%%%%%%%%%%%%%%%%%%%%%%%%%%%%%%%%%%
\begin{figure}[h]
  \begin{center}
  \begin{tabular}{c}
   \includegraphics[width=0.4\textwidth]{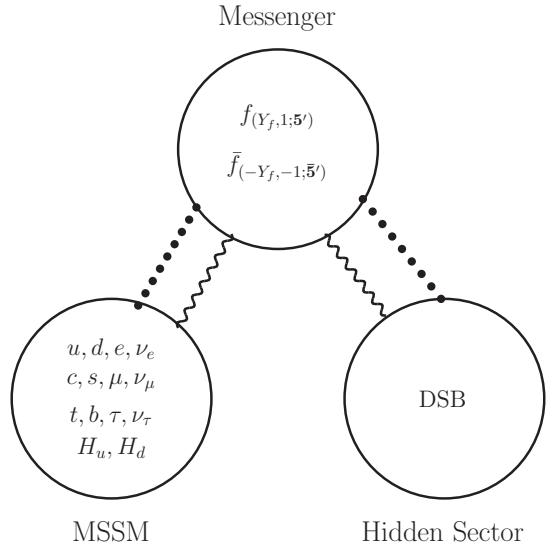}
   \end{tabular}
  \end{center}
  \caption{An effSUSY through \LSTMM.  }
\label{fig:LSTMM}
\end{figure}

%%%%%%%%%%%%%%%%%%%%%%%%%%%%%%%%%%%%%%%%%%%%%%%%%%%%%%%%%%%%%%%%%%%%%%%%%%%%%%%%%%%%%%%%%%%%
\section{String example}\label{sec:string}
%%%%%%%%%%%%%%%%%%%%%%%%%%%%%%%%%%%%%%%%%%%%%%%%%%%%%%%%%%%%%%%%%%%%

In this section, we discuss the \LSTZPM\ and \LSTMM\ based on the \Z$_{12-I}$ orbifold model of Ref. \cite{GMSBstSU3}, where the 4D gauge group is
\dis{
{\rm SU}&(3)_c\times{\rm SU(3)}_W\times{\rm U(1)}^3\times{\rm SU(2)}_n\\
&\times{\rm SU(5)}'\times{\rm SU(3)}'\times{\rm U(1)}^{'\,2}. \nonumber
}
The gauge groups SU(2)$_n$ and SU(3)$'$ are completely broken by the Higgs mechanism.
The gauge group SU(2)$_n$ is neutral, \ie it does not contribute to the SM hypercharge $Y$. But the hidden sector gauge group SU(3)$'$ contributes to $Y$: $\three'\to (\frac{-1}3~\frac{-1}3~\frac{2}3)$ and $\threeb'\to (\frac{1}3~\frac{1}3~\frac{-2}3)$.
To break SU(3)$'$ completely with these hypercharge contributions, we assign VEVs to two independent even $\Gamma$ fields in the lower box of Table \ref{table:Nsinglets}.
Thus at the GUT scale, SU(3)$_c\times$SU(3)$_W\times$U(1) is broken to SU(3)$_c\times$SU(2)$_W\times$U(1)$_Y$.  Then, at the electroweak scale we have the following gauge group
\dis{
&{\rm SU(3)}_c\times{\rm SU(2)}_W\times{\rm U(1)}_Y\times{\rm SU(5)}' \\
&\times{\rm U(1)}_1\times{\rm U(1)}_2\times{\rm U(1)}_3
\times{\rm U(1)}_4\times{\rm U(1)}_5
}
where the five additional U(1) charges are
\begin{align}
\begin{array}{l}
Q_1= (6~6~\textstyle{-6}~0~0~0~0~0)(0~0~0~0~0~0~0~0)' \\
Q_2=  (0~0~0~6~6~6~0~0)(0~0~0~0~0~0~0~0)' \\
Q_3=  (0~0~0~0~0~0~2~2)(0~0~0~0~0~0~0~0)'\\
Q_4=  (0~0~0~0~0~0~0~0)(4~4~4~4~4~0~0~0)'\\
Q_5= (0~0~0~0~0~0~0~0)(0~0~0~0~0~4~4~4)'\,.
\end{array}\label{U1s}
\end{align}
This model leads to one SU(5)$'$ family, breaking SUSY at an intermediate scale $\Lambda_h$, and the SM gauge group with three families and one pair of Higgs doublets. It contains the ingredients of the MSSM and the DSB source.

Supersymmetry breaking by one $\ten'$ and one $\fiveb'$ is achieved by the starred fields of Table \ref{table:Hidden}. A possible combination with  one $\ten'$ and one $\fiveb'$ is possible with the hidden sector gauginos \cite{Meurice84,KimNilles09}
\dis{
X' \propto \epsilon_{acfgh}\tilde{G'}^a_b\tilde{G'}^c_d \ten'^{eb}\fiveb'_e \ten'^{fd} \ten'^{gh}
\label{CompositeX}
}
which carries $Y'=Q_4+\frac15 Q_5=0$, but breaks its orthogonal combination U(1)$_{Y'_\perp}$. Also, the $Y$ value of $X'$ is also zero. Thus, the model realizes the scenarios discussed in Sec. \ref{sec:MedPhen}. The messenger sector charges determine whether it is \LSTZPM\ or \LSTMM.

%%%%%%%%%%%%%%%%%%%%%%%%%%%%%%%%%%%%%%%%%%%%%%%%%%%%%%%%%%%%%%%%%%%%
\subsection{Hidden sector SU(5)$'$, gauge mediation, messengers, and $R$-parity}
\label{subsec:SU(5)pr}

The hidden sector SU(5)$'$ representations are shown in Table \ref{table:Hidden}. Removing vectorlike pairs, we obtain one family SU(5)$'$ model below the GUT scale. The light \ten$'$ and $\fiveb'$ are marked with a star. These chiral fields carry the vanishing $Y$ charge, and SUSY breaking at the scale $\Lambda_h$ does not break U(1)$_Y$ of the SM.
%%%%%%%%%%%%%%%%%%%%%%%%%%%%%%%%%%%%%%%%%%%%%%%%%%%%%%%%%%%%%%%%%
 \begin{table}[t]
\begin{center}
\begin{tabular}{|c|c|c|}
\hline  $P+n[V\pm a]$ &  $\Gamma$ &
(Repts.)$_{Y[Q_1,Q_2,Q_3,Q_4,Q_5]}$
\\[0.2em]
\hline &&\\ [-1.1em]
$(\frac{1}6^2~\frac{-1}6~\frac{1}{6}^3~\frac{1}{4}^2)
 (\underline{\frac{-3}4~\frac{1}{4}^4}~\frac{-1}{4}^3)'_{T1_-}$ & $2$ & $
 (\one;\fiveb', \one)_{0~[3,3,1;1,-1]}^L$
\\[0.4em]
$(\frac{1}{6}^2~\frac{-1}{6}^4~ 0^2)
 (\underline{\frac{1}{2}~\frac{1}{2}~\frac{-1}{2}^3} ~\frac{-1}{6}^3)'_{T2_+}$
  & $-1$ & $\star~({\bf 1};\ten', \one)_{0~[3,-3,0;-2,-2]}^L$
\\[0.4em]
 $(0^6~\underline{\frac14~\frac{-3}{4}})
 (\underline{\frac34~\frac{-1}{4}^4~}~\frac{1}{4}^3)'_{T3}$
  & $-1$ & $(\two_n;{\bf 5}',\one)_{0~[0,0,-1;-1,3]}^L$
\\[0.4em]
  $(0^6~\underline{\frac34~\frac{-1}{4}})
 (\underline{\frac{-3}4~\frac{1}{4}^4}~\frac{-1}{4}^3)'_{T9}$
  & $1$ & $(\two_n;\fiveb',\one)_{0~[0,0,1;1,-3]}^L$
  \\[0.4em]
 $(0^3~\frac{-1}{3}^3~\frac{1}{4}~\frac{1}{4})
 (\underline{\frac{-3}4~\frac{1}{4}^4}~\frac{1}{12}^3)'_{T7_0}$
  & $-1$ & $\star~(\one;\fiveb',\one)_{0~[0,-6,1;1,1]}^L$
\\[0.4em]
 $(\frac{1}6^2~\frac{-1}6~ \frac{1}6^3~\frac{-1}{4}^2)
 (\underline{\frac{3}4 ~\frac{-1}{4}^4}~\frac{1}{4}^3)'_{T7_-}$
  & $0$ & $(\one;\five',\one)_{0~[3,3,-1;-1,3]}^L$
\\[0.4em]
$(0^6~\frac{-1}{2}~\frac{-1}{2})
 (\underline{\textstyle -1~0^4}~0^3)'_{T6}$ & $-2$ & $
 3\cdot(\one;\fiveb', \one)_{0~[0,0,-2;-4,0]}^L$
\\
$(0^6~\frac{-1}{2}~\frac{-1}{2})
 (\underline{\textstyle 1~0^4}~0^3)'_{T6}$ & $-2$ & $
 2\cdot(\one;{\bf 5}', \one)_{1~[0,0,-2;4,0]}^L$
\\
$(0^6~\frac{1}{2}~\frac{1}{2})
 (\underline{\textstyle -1~0^4}~0^3)'_{T6}$ & $2$ & $
 2\cdot(\one;\fiveb', \one)_{-1~[0,0,2;-4,0]}^L$
\\
$(0^6~\frac{1}{2}~\frac{1}{2})
 (\underline{\textstyle 1~0^4}~0^3)'_{T6}$ & $2$ & $
 3\cdot(\one;\five',\one)_{0~[0,0,2;4,0]}^L$
\\[0.2em]
\hline
\end{tabular}
\end{center}
\caption{Hidden sector SU(5)$'$  representations under
SU(2)$_n\times $SU(5)$'\times $SU(3)$'$. After removing vectorlike
representations by $\Gamma=$ even integer singlets, the starred
representations remain.} \label{table:Hidden}
\end{table}
%%%%%%%%%%%%%%%%%%%%%%%%%%%%%%%%%%%%%%%%%%%%%%%%%%%%%%%%%%%%%%%%%%%
As pointed out in Ref. \cite{SU5}, one family $\ten'+\fiveb'$ of a confining SU(5)$'$ breaks SUSY. For the \LSTZPM\ and \LSTMM, we require the confining scale $\Lambda_h$
below $10^{12}$ GeV \cite{GMSBrev,Kim07}.

Even though the singlet combination $\ten'\ten'\ten'\fiveb'$ is not
possible with one $\ten'$ and one $\fiveb'$, SUSY breaking can be parameterized by the hidden sector gauge field strength ${\cal W'}^\alpha{\cal W}'_\alpha$ \cite{Murayama07,GMSBstSU3}, for the messenger $f$ and $\bar{f}$,
\begin{equation}
{\cal L}=\int d^2\theta \left[\xi(\cdots) f\bar{f}
 {\cal W'}^\alpha{\cal W'}_\alpha+\eta(\cdots)
 f\overline{f}\right]+{\rm h.c.}\label{yukawast}
\end{equation}
where we have in general the holomorphic functions $\xi$ and
$\eta$ of singlet chiral fields, $S_1, S_2,\cdots$. Assuming the singlet VEVs at the string scale, Ref. \cite{GMSBstSU3} discussed the GMSB. On the other hand, it is generally expected that some of singlet VEVs are smaller than the string scale. Then, $f$ and $\bar{f}$ carrying SU(3)$_c$ and SU(2)$_W$ charges have negligible couplings to ${\cal W'}^\alpha{\cal W'}_\alpha$, and the original GMSB \cite{DineNeslson} is probably not realized in the model of Ref. \cite{GMSBstSU3}. The main reason is that SU(5)$'$-colored $f$ and $\bar{f}$ are multiplied to ${\cal W'}^\alpha{\cal W'}_\alpha$ together with many small VEV singlet fields. We argue that this may be of general nature. This leads us to consider \LSTZPM\ and \LSTMM\ discussed in Sec. \ref{sec:MedPhen}. For simplicity, we choose only one relatively light (at the $10^{13-15}$ GeV) pair of $f$ and $\bar{f}$ from Table \ref{table:Hidden}, and assume all the other $\five'$ and $\fiveb'$ are sufficiently heavy such that the consideration of one pair of $f$ and $\bar{f}$ is sufficient.

 %%%%%%%%%%%%%%%%%%%%%%%%%%%%%%%%%%%%%%%%%%%%%%%%%%%%%%%%%%%%%%%%%%%%
 \begin{table}[t]
\begin{center}
\begin{tabular}{|c|c|c|c|c|}
\hline  Sector  &
(Repts.)$_{Y[Q_1,Q_2,Q_3,Q_4,Q_5]}$
 &$\Gamma$ &Label\\[0.2em]
\hline &&&\\[-1.3em]
 $T_{4_-}$  & $ 3\cdot(\three,\two)_{1/6~[0,0,0;0,0]}^L$
  & $1$ &$ q_1,~ q_2,~  q_3$
\\[0.2em]
$T_{4_-}$   &$2\cdot(\threeb,\one)_{-2/3~[-3,3,2;0,0]}^L$& $3$ &$ u^c,~  c^c$
\\[0.2em]
$T_{7_+}$   &$(\threeb,\one)_{-2/3~[0,6,-1;5,1]}^L$& $1$ &$ t^c$
\\[0.2em]
$T_{2_0}$   &$(\threeb,\one)_{1/3~[3,-3,0;0,-4]}^L$& $-1$ &$ b^c$
\\[0.2em]
$T_{4_-}$  &$2\cdot(\threeb,\one)_{1/3~[-3,3,-2;0,0]}^L$ & $1$ &$ d^c,~ s^c$
\\[0.4em]
\hline &&&\\[-1.3em]
 $T_{4_-}$  &$ 3\cdot(\one,\two)_{-1/2~[-6,6,0;0,0]}^L$  & $1$ &$l_1,l_2,l_3$
\\[0.2em]
$U_1$ & $\one_{1~[9,3,-2;0,0]}^L $  &$1$& $e^c$
\\[0.2em]
$T_{3}$&  $\one_{1~[0,-6,1;5,-3]}$ & $1$ &$ \mu^c$
\\[0.2em]
$T_{1_0}$ & $ \one_{1~[0,6,-1;5,1]}^L$  &  $3$ &$\tau^c$
\\[0.2em] \hline &&&\\[-1.1em]
 $T_{1_0}$  &$ (\one,\two)_{1/2~[0,6,-1;5,1]}^L$ & $0$ &$H_u$
\\[0.2em]
$T_{7_+}$  &$(\one,\two)_{-1/2~[-6,0,-1;5,1]}^L$ & $-2$ &$H_d$
\\[0.2em]\hline
\end{tabular}
\end{center}
\caption{Three families of quarks and leptons and a pair of Higgs doublets of \cite{GMSBstSU3}. The quark singlets $d^c$ and $b^c$ are interchanged from those of Ref. \cite{GMSBstSU3} to have an effSUSY. The lepton singlets are taken from Table \ref{table:Csinglets} in Appendix B.} \label{table:Families}
\end{table}
%%%%%%%%%%%%%%%%%%%%%%%%%%%%%%%%%%%%%%%%%%%%%%%%%%%%%%%%%%%%%%%%%%%

For \LSTZPM, we can choose for example $f=\five'_0$ from the $T_3$ sector and $\bar f=\fiveb'_0$ from the $T_9$ sector.
For \LSTMM, we must choose $f=\five'_1$ from $T_6$ and $\bar f=\fiveb'_{-1}$ from $T_6$. Therefore, the model presented in \cite{GMSBstSU3} has the basic ingredients for \LSTZPM\ and \LSTMM.

As discussed in \cite{GMSBstSU3}, there appear three SM family members of Table \ref{table:Families}. Also, only one pair of Higgs doublets results because the superpotential for three SU(3)$_W$ anti-triplets must be symmetric under exchange of two superfields. But, SU(3)$_W$ invariance needs an antisymmetric SU(3)$_W$ indices, needing an antisymmetric flavor indices. This leads to one pair of massless Higgsinos naturally. So by SUSY, we have a pair of massless Higgs doublets at the SU(3)$_W$ breaking scale(the GUT scale).
The TeV scale $\mu$ is generated by norenormalizable superpotential terms \cite{Mu}, which will be worked out explicitly in the present string model \cite{KimKyae11}.
This fulfils all the requirements of \LSTZPM\ and \LSTMM.

The $Y'$ quantum number is
\dis{
 Y'=Q_3+\frac15 Q_4.
}
From Table \ref{table:Families}, we find that $ \tilde{t}_L, \tilde{t}_R, \tilde{b}_L, \tilde{b}_R, \tilde{\tau}_L, \tilde{\tau}_R, \tilde{\nu}_{\tau}, H_u$, and $H_d$ carry the vanishing $Y'$. Also, $X'$ of Eq. (\ref{CompositeX}) carries the vanishing $Y'$. These provide the needed quantum numbers of Fig. \ref{fig:LSTZPM} and Table \ref{table:BminusL12}. The string model \cite{GMSBstSU3} is a kind of the flavor unification model, and different families need not have the same $Y'$ quantum numbers. In GUTs descending from E$_6$ which is not a flavor unification model, the family distinction of $Y'$ is not present and hence there is the problem of low energy baryonic or leptophobic U(1)$_{Y'}$ \cite{GUTsE6}. The GUTs from F-theory construction \cite{KimFths} is not free from the low energy U(1)$_{Y'}$ problems.

The proton longevity is the key requirement in the SUSY extension of the SM, usually achieved in terms of the $R$-parity. This is a parity where the SM matter superfields carry the odd parity while the Higgs superfields carry the even parity. In the orbifold compactification, one combinations, say U(1)$_\Gamma$, of U(1)s is the covering gauge symmetry of the $R$-parity. If some even $\Gamma$ scalars, with the smallest $|\Gamma|$ normalized as 1, develop VEVs, then we obtain the $R$-parity \cite{KimKyaeSM,GUT07}:
\begin{equation}
{\textrm U(1)}_\Gamma\to P.\label{Rparity}
\end{equation}
If some odd $\Gamma$ scalars develop VEVs also, then the $R$-parity is spontaneously broken.
To have an $R$-parity, we define the following $\Gamma$,
\begin{equation}
\Gamma=(0~0~0~3~3~0~2~2)(0^8)\label{UGamma}
\end{equation}
We require that only the even $\Gamma$ fields are allowed to develop VEVs. If we used some global symmetries in string models, they must be approximate \cite{apprPQ,apprR}. The $\Z_4$ and other discrete $R$ symmetries from an approximate global U(1)$_R$ have been tabulated recently, where discrete anomaly-free conditions have been imposed in addition \cite{Z4}.

%%%%%%%%%%%%%%%%%%%%%%%%%%%%%%%%%%%%%%%%%%%%%%%%%%%%%%%%%%%%%%%%%%%%%%%%%%%%%%%%%%%%%%%%%
\subsection{VEVs leading to one $Z'$}
\label{subsec:oneZpr}

There are five extra U(1)s, Eq. (\ref{U1s}), beyond U(1)$_Y$ of the SM. To have one light $Z'$, we need four independent singlet VEVs. Three of these are provided by the starred singlet fields of Table \ref{table:Nsinglets}:
\dis{
&S'_1: U_{2}(6\star)[0,12,0,0,0]\\
&S'_2: T_{2_0}(2\star)[0,6,0,0,-4]\\
&S'_3: T_{4_+}(-2\star)[6,-6,0,0,-4],
}
where the sector, $\Gamma$, and five U(1)$'$ quantum numbers are shown. The fourth singlet combination is the quantum number of $X'$ of Eq. (\ref{CompositeX})
\dis{
X': ~ [0,0,1;-5,0]
}
(as shown in \cite{Meurice84,KimNilles09}) which carries $Q_3=1$ and $Q_4=-5$ and hence carries $Y'=0$. The other combination orthogonal to $Y'$, say $Y'_\perp$, is broken by this dynamical composite, and we obtain one light $Z'$ model.

The hierarchy of VEVs is that $\langle S'_1\rangle, \langle S'_2\rangle$, and $ \langle S'_3\rangle$ is much greater than \MGUT\ and the U(1)$'_\perp$ breaking scale is at the \LSTZPM\ scale, \ie the hidden sector scale $\Lambda_h \lesssim 10^{12}$ GeV. Therefore, below \MGUT\ we may consider two light $Z'$s: $Z'$ and $Z'_\perp$.

\begin{figure}[t]
  \begin{center}
  \begin{tabular}{c}
   \includegraphics[width=0.35\textwidth]{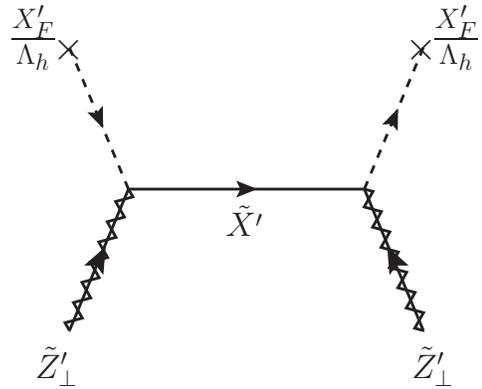}
   \end{tabular}
  \end{center}
  \caption{The $\tilde{Z}'_\perp$ mass splitting via the SUSY breaking through $X'$. The $\tilde{Z}'_\perp$ line is sawed. The dimensional parameter is the confining scale $\Lambda_h$.  Even though the $\tilde{Z}'_\perp$
  mass is smaller than that of $\tilde{Z}'$, the mass splitting of the $\tilde{Z}'_\perp$ multiplet is larger than that of $\tilde{Z}'$. }
\label{fig:Zpperpino}
\end{figure}

Note that both the $Z'_\perp$ and $\tilde{Z}'_\perp$ mass scales are $\Lambda_h$ because the SUSY breaking $F$-term carries a nonvanishing $Y'_\perp$ charge also. Compared to the other U(1)-priminos, there are two relatively light inos, $\tilde Z'$ and $\tilde Z'_\perp$. Among these,  the mass splitting of the $Z'_\perp$ multiplet is greater than that of the  $Z'$ multiplet. The mass splitting of the $\tilde{Z}'_\perp$ multiplet is of order $\Lambda_h$ as shown in Fig. \ref{fig:Zpperpino} while the mass splitting of the $\tilde Z'$ multiplet is of order $\Lambda_h^2/M_{\rm mess}$ as shown in Fig. \ref{fig:ZpMass}.  Thus, we can consider only a light $\tilde Z'$ mass splitting for the light superpartners as discussed in Sec. \ref{sec:MedPhen}.\\

%%%%%%%%%%%%%%%%%%%%%%%%%%%%%%%%%%%%%%%%%%%%%%%%%%%%%%%%%%%%

The breaking scale of U(1)$_{Y'}$ or the $Z'$ mass is required to be somewhat below the GUT scale, and it is not required for it to be less than $\Lambda_h$. The only requirement is that the SUSY breaking scale via $Z'$ mediation, \ie the supertrace of the $Z'$ SUSY sector or the Zprimino mass, is of order a TeV scale.

%%%%%%%%%%%%%%%%%%%%%%%%%%%%%%%%%%%%%%%%%%%%%%%%%%%%%%%%%%%%%%%%%%%%%%%%
%%%%%%%%%%%%%%%%%%%%%%%%%%%%%%%%%%%%%%%%%%%%%%%%%%%%%%%%%%%%%%%%%%%%%%%%
\section{Conclusion}\label{sec:Conclusion}

Motivated by the existing string compactification model, we investigated the possibility of the U(1)$'$ contribution to the mediation mechanism, leading to an effective SUSY.  The first two family sfermions are required to be above 100 TeV but the third family fermions and the Higgs doublets are in the TeV region. For a few parameter ranges, we calculated the spectra of superpartners in the \LSTZPM\ and MM. In the \LSTZPM\  scenario, the Higgs fields survive down to the electroweak scale by tuning the ratio of the DSB scale $\Lambda_h$ and the messenger scale $M_{\rm mess}$. In the mixed mediation scenario, it is shown that an additional fine tuning between parameters of the Higgs boson mass matrix is required as in Ref. \cite{Yavin07}. We noted that the Zprimino needs not be at a TeV scale. It is required that it is somewhat lighter than the messenger scale. We also discussed the needed conditions among the fields obtained in the string construction of Ref. \cite{GMSBstSU3} for the \LSTZPM\ or the \LSTMM.

%%%%%%%%%%%%%%%%%%%%%%%%%%%%%%%%%%%%%%%%%%%%%%%%%%%%%%%%%%%%%%%%%%%%%%%%%%%%
\vskip 0.2cm
\noindent {\bf Acknowledgments}: {This work is supported in part by the National Research Foundation  (NRF) grant funded by the Korean Government (MEST) (No. 2005-0093841),
and KSJ is supported in addition by the JSPS Grant-in-Aid 21-09224 for JSPS Fellows.
}

%%%%%%%%%%%%%%%%%%%%%%%%%%%%%%%%%%%%%%%%%%%%%%%%%%%%%%%%%%%%%%%%%%%%%%%%%%%%

%\newpage
%%%%%%%%%%%%%%%%%%%%%%%%%%%%%%%%%%%%%%%%%%%%%%%%%%%%%%%%%%%%%%
\vskip 0.5cm
\centerline{\bf Appendix A: Soft terms in U(1)$'$ mediation}
\vskip 0.3cm

In this Appendix, we present the RG equations for soft terms in
the Zprimino mediation.
We consider the case that the U(1)$^\prime$ breaking scale is much higher than
$M_{\tilde Z^\prime}$, for which the U(1)$^\prime$ vector superfield acquires
a large supersymmetric mass.

At the messenger scale $M_{\rm mess}$, the Zprimino acquires soft
mass at the one-loop level while the MSSM gaugino masses vanish:
\dis{
&M_{\tilde Z^\prime} = -\frac{g^2_{Y^\prime}}{8\pi^2}\frac{F_{\rm mess}}{M_{\rm mess}}, \\
&M_a = 0.
}
The RG equation for gaugino masses is written as
\dis{
&\mu \frac{d}{d \mu}\left(\frac{M_{\tilde Z^\prime}}{g^2_{Y^\prime}}\right) =0, \\
&\mu \frac{d}{d \mu}\left(\frac{M_a}{g^2_a}\right) =
\frac{c_a}{8\pi^2 b_{Y^\prime}}\mu\frac{d M_{\tilde Z^\prime}}{d\mu},
}
for $M^\prime<\mu<M_{\rm mess}$.
Here $b_{Y^\prime}=\sum_i Y^{\prime 2}_i$ is the beta function coefficient for
$U(1)^\prime$, and $c_a$ are given by
\dis{
&c_Y=\sum \Big[\,6\left(\frac16\right)^2Y^{'2}_{Q}+ 3\left(\frac13\right)^2Y^{\prime 2}_{U^c}
\\
&\hspace{1.6cm}
+  3\left(\frac13\right)^2Y^{\prime 2}_{D^c} + 2\left(\frac12\right)^2 Y^{\prime 2}_L
+ Y^{\prime 2}_{E^c} \Big], \\
&c_2= \sum\Big[\,3Y^{'2}_{Q}  +Y^{\prime 2}_L\Big],\\
&c_3= \sum\Big[\,2Y^{'2}_{Q}  +Y^{'2}_{U^c} +Y^{'2}_{D^c}\Big],
}
where the sum is over U(1)$^\prime$-charged families.
The U(1)$^\prime$ vector multiplet decouples in a supersymmetric way at energy
scales below $M^\prime$, which is assumed to be much higher than
$M_{\tilde Z^\prime}$. Hence, at $\mu<M^\prime$, the MSSM gaugino masses are determined by
\dis{
\mu\frac{d}{d \mu}\left(\frac{M_a}{g^2_a}\right) =0.
}
The RG equations for the soft terms associated with the third family sfermions
and Higgs bosons are the same as the MSSM at energy scales below $M_{\rm mess}$.
For the first two family sfermions, one finds
\dis{
\mu\frac{d m^2_i}{d\mu} = -\frac{Y^{\prime 2}_i}{2\pi^2}g^2_{Y^\prime}
M^2_{\tilde Z^\prime},
}
because they couple to the U(1)$^\prime$ vector multiplet, and have negligible
Yukawa couplings.

%%%%%%%%%%%%%%%%%%%%%%%%%%%%%%%%%%%%%%%%%%%%%%%%%%%%%%%%%%%%%%
\vskip 0.5cm
\centerline{\bf Appendix B: Singlets}
\vskip 0.3cm

In this Appendix, we list all charged singlets in Table \ref{table:Csinglets} and all neutral singlets in Table \ref{table:Nsinglets}, which were needed in Sec. \ref{sec:string} but not listed in Ref. \cite{GMSBstSU3}. The shift vector and the Wilson line are

\begin{align}
 V&=\textstyle
 (\frac12~\frac12~\frac12~\frac{1}{6}~\frac{1}{6}~\frac{1}{6}~\frac{1}{4}
 ~\frac{1}{4})(\frac{1}{4}~\frac{1}{4}~\frac{1}{4}~\frac{1}{4}~\frac{1}{4}
 ~\frac{1}{12}~\frac{1}{12}~\frac{1}{12})'
 \label{Z12ImodelC}\\
  a_3&=\textstyle
(\frac{1}{3}~\frac{1}{3}~\frac{2}{3}~ 0~0~0~0~0)(
 0~0~0~0~0~\frac{1}{3}~\frac{1}{3}~\frac{-2}{3})'.\label{WilsonLine}
\end{align}

\begin{widetext}

%%%%%%%%%%%%%%%%%%%%%%%%%%%%%%%%%%%%%%%%%%%%%%%%%%%%%%%%%%%%%%%%%
 \begin{table}[b]
\begin{center}
\begin{tabular}{|c|c|c|}
\hline  $P+n[V\pm a]$ &  $\Gamma$ &
No.$\times$(Repts.)$_{Y[Q_1,Q_2,Q_3,Q_4,Q_5]}$
\\
\hline &&\\[-1.1em]
$({\frac{1}2~\frac{1}2~\frac{-1}2~\frac{1}{2}~\frac{1}{2}~
 \frac{-1}{2}~\frac{-1}{2}~\frac{-1}{2}}) (0^8)'_{U_1}$ &  $1$ & $
 \one_{1~[9,3,-2;0,0]}^L\to e^c$
\\

$(0~0~0~0~0~\textstyle{-1}~\underline{
\textstyle{-1}~0}) (0^8)'_{U_3}$ &  $-2$ & $ 2\cdot\one_{1~[0,-6,-2;0,0]}^L$
\\
$(0~0~0~\frac23~\frac23~\frac{-1}3~\frac{-1}4~\frac{-1}4)
 (\frac{1}4~\frac{1}4~\frac{1}4~\frac{1}4~\frac{1}4~\frac{1}{12}~\frac{1}{12}~\frac{1}{12}
 )'_{T_{1_0}}$ &  $3$  & $ \one_{1~[0,6,-1;5,1]}^L\to\tau^c$
\\
$(0~0~0~\frac{-1}3~\frac{-1}3~\frac{-1}3~\frac{-1}4~\frac{-1}4)
 (\frac{1}4~\frac{1}4~\frac{1}4~\frac{1}4~\frac{1}4~\frac{1}{12}~\frac{1}{12}~\frac{1}{12}
 )'_{T_{1_0}}$ &  $-3$ & $2\cdot \one_{1~[0,-6,-1;5,1]}^L$
\\
$(\frac{1}{6}~\frac{1}{6}~\frac{-1}{6}~\frac{1}{6}~\frac{1}{6}~\frac{1}{6}~
\frac{1}4~\frac{1}4)
 (\frac{-1}4~\frac{-1}4~\frac{-1}4~\frac{-1}4~\frac{-1}4~\frac{1}4~\frac{1}4~\frac{1}4~
 )'_{T_{1_-}}$ &  $1$ & $2\cdot \one_{-1~[3,3,1;-5,3]}^L$
\\
$(0~0~0~0~ 0~\textrm{-1}~\frac{1}4~\frac{1}4)
 (\frac{1}4~\frac{1}4~\frac{1}4 ~\frac{1}4~\frac{1}4~\frac{-1}4~\frac{-1}4~\frac{-1}4)'_{T_{3}}$
 &  $1$ & $(2_L+1_R)\one_{1~[0,-6,1;5,-3]}\to \mu^c$
\\
$(0~0~0~0~ 0~0~\underline{\frac{1}4~\frac{-3}4})
 (\frac{1}4~\frac{1}4~\frac{1}4 ~\frac{1}4~\frac{1}4~\frac{-1}4~\frac{-1}4~\frac{-1}4)'_{T_{3}}$
 &  $-1$ & $ (6_L+6_R)\one_{1~[0,0,-1;5,-3]}$
\\
$(\frac{1}6~\frac{1}6~\frac{-1}6~\frac{1}6~\frac{1}6~\frac{-5}6~
\frac{-1}2~\frac{-1}2) (0^8)'_{T_{4_-}}$ & 0& $2\cdot\one_{1~[3,-3,-2;0,0]}^L$
\\
$(\frac{1}6~\frac{1}6~\frac{-1}6~\frac{1}6~\frac{1}6~\frac{1}6~
\underline{\frac{-1}2~\frac{-1}2}) (0^8)'_{T_{4_-}}$ & 0& $
 12_L\cdot\one_{1~[3,3,0;0,0]}$
\\
$(\frac{-1}3~\frac{-1}3~\frac{1}3~\frac{-1}3~\frac{-1}3~\frac{-1}3~0~0)
 (0^8)'_{T_{4_-}}$ & $-2$ & $ 7_R\cdot\one_{1~[-6,-6,0;0,0]}$
\\
$(\frac{-1}3~\frac{-1}3~\frac{1}3~\frac{2}3~\frac{2}3~\frac{-1}3~0~0)
 (0^8)'_{T_{4_-}}$ & $4$ & $ 3_R\cdot\one_{1~[-6,6,0;0,0]}$
\\
$(\frac{1}6~\frac{1}6~\frac{-1}6~\frac{1}6~\frac{1}6~\frac{-5}6~\frac{-1}2~\frac{-1}2)
 (0^8)'_{T_{4_-}}$ & $-4$ & $ 2_R\cdot\one_{1~[3,-3,-2;0,0]}$
\\

$(\frac{1}3~\frac{1}3~\frac{-1}3~\frac{-1}3~\frac{-1}3~
\frac{2}3~\frac{1}4~\frac{1}4)
 (\frac{-1}4~\frac{-1}4~\frac{-1}4~\frac{-1}4~\frac{-1}4~\frac{-1}{12}
 ~\frac{-1}{12}~\frac{-1}{12})'_{T_{7_+}}$ &  $-1$ & $\one_{-1~[6,0,1;-5,-1]}^L$
\\
$(\frac{-1}6~\frac{-1}6~\frac{1}6~\frac{1}6~\frac{1}6~
\frac{1}6~\underline{\frac{3}4~\frac{-1}4})
 (\frac{-1}4~\frac{-1}4~\frac{-1}4~\frac{-1}4~\frac{-1}4~\frac{-1}{12}
 ~\frac{-1}{12}~\frac{-1}{12})'_{T_{7_+}}$ &  $2$ & $2\cdot\one_{-1~[-3,3,1;-5,-1]}^L$
\\
$(\frac{1}6~\frac{1}6~\frac{-1}6~\frac{1}6~\frac{1}6~
\frac{1}6~\frac{-1}4~\frac{-1}4)
 (\frac{1}4~\frac{1}4~\frac{1}4~\frac{1}4~\frac{1}4~\frac{-1}4~\frac{-1}4~
 \frac{-1}4)'_{T_{7_-}}$& 0 & $2\cdot\one_{1~[3,3,-1;5,-3]}^L$
\\
[0.4em] \hline &&\\[-1.1em]
$(\frac{1}6~\frac{1}6~\frac{-1}6~\frac{1}6~\frac{1}6~\frac{1}6~\frac{1}4~\frac{1}4)
 (\frac{1}4~\frac{1}4~\frac{1}4~\frac{1}4~\frac{1}4~\frac{3}4~\frac{-1}4~\frac{-1}4)'_{T_{1_-}}$
 &  $2$ & $\one_{1~[3,3,1;5,1]}^L$
\\
$(\frac{1}6~\frac{1}6~\frac{-1}6~\frac{-1}6~\frac{-1}6~\frac{-1}6~0~0)
 (0^5~\frac{1}3~\frac{-2}3~\frac{-2}3)'_{T_{2_+}}$ & $-1$  & $
  \one_{1~[3,-3,0;0,-4]}^L$
\\
$(0~0~0~\frac{-1}3~\frac{-1}3~\frac{2}3~0~0)
 (0~0~0~0~0~\frac{-2}3~\frac{1}3~\frac{1}3)'_{T_{4_0}}$
 &  $-2$ & $3\cdot \one_{-1~[0,0,0;0,0]}^L$
\\
 $(\frac{1}{3}~\frac{1}{3}~\frac{-1}{3}~\frac{-1}{3}~\frac{-1}{3}~\frac{-1}{3}~0~0)
 (0^5~\frac{2}{3}~\frac{-1}{3}~\frac{-1}{3}~)'_{T_{4_+}}$
 &  $-2$ & $ 3\cdot \one_{1~[6,-6,0;0,0]}^L$
\\
 $(\frac{-1}{3}~\frac{-1}{3}~\frac{1}{3}~\frac{1}{3}~\frac{1}{3}~\frac{1}{3}~\frac{-1}{2}~
 \frac{-1}{2}) (0^5~\frac{2}{3}~\frac{-1}{3}~
 \frac{-1}{3})'_{T_{4_+}}$
 &0 & $ 2\cdot \one_{1~[-6,6,-1;0,0]}^L$
\\
$(\frac{-1}6~\frac{-1}6~\frac{1}6~\frac{1}6~\frac{1}6~\frac{1}6~\frac{1}2~\frac{1}2)
(0^5~ \frac{-1}{3}\underline{\frac{2}{3}~\frac{-1}{3}})'_{T_{4_+}}$
 &  $3$ & $4\cdot \one_{-1~[-3,3,2;0,0]}^L$
\\
$(0~0~0~\frac{-1}3~\frac{-1}3~\frac{-1}3~\frac{1}4~\frac{1}4)
 (\frac{-1}4~\frac{-1}4~\frac{-1}4~\frac{-1}4~\frac{-1}4~ \frac{-5}{12}~\underline{\frac{-5}{12}~\frac{7}{12}})'_{T_{7_0}}$
 & $-1$  & $2\cdot \one_{-1~[0,-6,1;-5,-1]}^L$
\\
$(\frac{1}6~\frac{1}6~\frac{-1}6~\frac{1}6~\frac{1}6~
\frac{1}6~\frac{-1}4~\frac{-1}4)
 (\frac{-1}4~\frac{-1}4~\frac{-1}4~\frac{-1}4~\frac{-1}4~\frac{-3}4~
 \frac{1}4~\frac{1}4)'_{T_{7_-}}$
 & 0& $ \one_{-1~[3,3,-1;-5,-1]}^L$
\\[0.4em]
\hline
\end{tabular}
\end{center}
\caption{The charged singlets. The fields in the lower box get $Y$ contribution from the SU(3)$'$ generators, and the underline means permutations.} \label{table:Csinglets}
\end{table}
%%%%%%%%%%%%%%%%%%%%%%%%%%%%%%%%%%%%%%%%%%%%%%%%%%%%%%%%%%%%%%%%%%%

%%%%%%%%%%%%%%%%%%%%%%%%%%%%%%%%%%%%%%%%%%%%%%%%%%%%%%%%%%%%%%%%%
 \begin{table}[t]
\begin{center}
\begin{tabular}{|c|c|c|}
\hline  $P+n[V\pm a]$ &  $\Gamma$ &
No.$\times$(Repts.)$_{Y[Q_1,Q_2,Q_3,Q_4,Q_5]}$
\\
\hline && \\ [-1.1em]
$(0~0~0~0~0~-1~\underline{1~0}) (0^8)'_{U_1}$ &$2\star$ & $2\cdot \one_{0~[0,-6,2;0,0]}^L$
\\
$({0~0~0~1~1~ 0~0~0}) (0^8)'_{U_2}$ & $ 6\star$ & $ \one_{0~[0,12,0;0,0]}^L\equiv S'_1$
\\
$(0~0~0~\frac{-1}3~\frac{-1}3~\frac{-1}3~\frac{-1}4~\frac{-1}4)
 (\frac{-1}4~\frac{-1}4~\frac{-1}4~\frac{-1}4~\frac{-1}4~
 \frac{-5}{12}~\frac{-5}{12}~\frac{-5}{12})'_{T_{1_0}}$
 & $ -3$  & $\one_{0~[0,-6,-1;-5,-5]}^L$
\\
$(\frac{-1}2~\frac{-1}2~\frac{1}2~\frac{1}6~\frac{1}6~\frac{1}6~
\frac{1}4~\frac{1}4)
 (\frac{1}4~\frac{1}4~\frac{1}4~\frac{1}4~\frac{1}4~\frac{1}{12}~\frac{1}{12}~\frac{1}{12}
 )'_{T_{1_0}}$ &$2\star$& $ \one_{0~[-9,3,1;5,1]}^L$
\\
$(\frac{1}6~\frac{1}6~\frac{-1}6~\frac{1}6~\frac{1}6~\frac{1}6~\frac{1}4~\frac{1}4)
 (\frac{1}4~\frac{1}4~\frac{1}4~\frac{1}4~\frac{1}4~\underline{\frac{3}{4}~\frac{-1}{4}}~
 \frac{-1}{4} )'_{T_{1_-}}$ &  $ 2\star$ & $ 2\cdot\one_{0~[3,3,1;5,1]}^L$
\\
$(0~0~0~\frac13~\frac13~\frac{-2}3~\frac{1}2~\frac{1}2)(0^5~\frac{-1}3~
\frac{-1}3~\frac{-1}3 )'_{T_{2_0}}$ &  $ 2\star$ & $ \one_{0~[0,0,2;0,-4]}^L$
 \\
$(\frac{-1}2~\frac{-1}2~\frac{1}2~\frac{-1}6~\frac{-1}6~\frac{-1}6~0~0)(0^5~\frac{-1}3~
\frac{-1}3~\frac{-1}3 )'_{T_{2_0}}$ &  $-1$ & $ 2\cdot\one_{0~[-9,-3,0;0,-4]}^L$
\\
$(0~0~0~\frac{1}3~\frac{1}3~\frac{1}3~\underline{\frac12~\frac{-1}2})(0^5~\frac{-1}3~
\frac{-1}3~\frac{-1}3 )'_{T_{2_0}}$ &  $ 2\star$ & $ 2\cdot\one_{0~[0,6,0;0,-4]}^L \equiv S'_2$
\\
$(\frac{-1}3~\frac{-1}3~\frac{1}3~\frac{1}3~\frac{1}3~\frac{1}3~\frac{-1}2~\frac{-1}2)(0^5~
\frac{1}3~\frac{1}3~\frac{1}3 )'_{T_{2_+}}$ & $0\star$ & $ \one_{0~[-6,6,-2;0,4]}^L$
\\
$(\frac{1}6~\frac{1}6~\frac{-1}6~\frac{-1}6~\frac{-1}6~\frac{-1}6~0~0)(0^5~
\frac{1}3~\frac{1}3~\frac{1}3 )'_{T_{2_+}}$ &  $-1$ & $ \one_{0~[3,-3,0;0,4]}^L$
\\
$(0~0~0~0~ 0~\textrm{1}~\frac{1}4~\frac{1}4) (\frac{1}4~\frac{1}4~\frac{1}4
 ~\frac{1}4~\frac{1}4~\frac{-1}4~\frac{-1}4~\frac{-1}4)'_{T_{3}}$ &1  & $
 (1_L+2_R)\one_{0~[0,6,1;5,-3]}^L$
\\
$(\frac{-1}6~\frac{-1}6~\frac{1}6~\frac{1}6~\frac{1}6~
\frac{-5}6~\frac{-1}4~\frac{-1}4)
 (\frac{-1}4~\frac{-1}4~\frac{-1}4~\frac{-1}4~\frac{-1}4~\frac{-1}{12}
 ~\frac{-1}{12}~\frac{-1}{12})'_{T_{7_+}}$ & $0\star$  & $ \one_{0~[-3,-3,-1;-5,-1]}^L$
\\[0.4em]
\hline &&\\[-1.1em]
$(\frac{1}6~\frac{1}6~\frac{-1}6~\frac{1}6~\frac{1}6~\frac{1}6~\frac{1}4~\frac{1}4)
 (\frac{1}{4}~\frac{1}4~\frac{1}4~\frac{1}4~\frac{1}4~\underline{\frac{3}4~
\frac{-1}4}~\frac{-1}4)'_{T_{1_-}}$ &  $ 2\star$ & $ 2\cdot \one_{0~[3,3,1;5,1]}^L$
\\
$(\frac{1}6~\frac{1}6~\frac{-1}6\frac{-1}6\frac{-1}6\frac{-1}6~0~0)
 (0^5~\underline{\frac{1}3~\frac{-2}3}~\frac{-2}3)'_{T_{2_+}}$ &  $-1$ & $
2\cdot \one_{0~[3,-3,0;0,-4]}^L$
\\
$(0~0~0~\frac{-1}3~\frac{-1}3~\frac{2}3 ~0~0)
 (0^5~\underline{\frac{1}3~\frac{-2}3}~\frac{1}3)'_{T_{4_0}}$ & $ -2\star$  & $
 6\cdot\one_{0~[0,0,0;0,0]}^L\equiv S'_3$
\\
 $(\frac{1}3~\frac{1}3~\frac{-1}3 \frac{-1}3~\frac{-1}3~\frac{-1}3 ~0~0)
 (0^5~\underline{\frac{2}{3}~\frac{-1}{3}}~\frac{-1}3)'_{T_{4_+}}$ &  $ -2\star$ & $
 6\cdot\one_{0~[6,-6,0;0,0]}^L$
\\
 $(\frac{-1}6~\frac{-1}6~\frac{1}6~\frac{1}6~\frac{1}6~\frac{1}6~\frac{1}2~\frac{1}2)
 (0^5~\underline{\frac{2}{3}~\frac{-1}{3}}~\frac{-1}3)'_{T_{4_+}}$ &  $3$ & $
 4\cdot\one_{0~[-3,3,2;0,0]}^L$
\\
 $(\frac{-1}6~\frac{-1}6~\frac{1}6~\frac{1}6~\frac{1}6~\frac{1}6~\frac{-1}2~\frac{-1}2)
 (0^5~\frac{-1}{3}~\frac{-1}{3}~\frac{2}3)'_{T_{4_+}}$ &$-1$ & $2\cdot\one_{0~[-3,3,-2;0,0]}^L$
\\
$(0~0~0~\frac{-1}3~\frac{-1}3~\frac{-1}3 ~\frac{1}4~\frac{1}4)
 (\frac{-1}4~\frac{-1}4~\frac{-1}4~\frac{-1}4~\frac{-1}4~ \frac{-5}{12}~\frac{-5}{12}~ \frac{7}{12})'_{T_{7_0}}$ &$-1$  & $
\one_{0~[0,-6,1;-5,-1]}^L$
\\
$(\frac{1}6~\frac{1}6~\frac{-1}6~\frac{1}6~\frac{1}6~\frac{1}6~\frac{-1}4~\frac{-1}4)
 (\frac{-1}4~\frac{-1}4~\frac{-1}4~\frac{-1}4~\frac{-1}4~
 \underline{\frac{1}{4}~\frac{-3}{4}}~\frac{1}4)'_{T_{7_-}}$& $0\star$  & $
 2\cdot\one_{0~[3,3,-1;-5,-1]}^L$
\\ [0.4em]
\hline
\end{tabular}
\end{center}
\caption{The same as Table \ref{table:Csinglets}, but for the neutral singlets. The starred even
$\Gamma$ quantum number fields can develop VEVs without breaking the $R$-parity.} \label{table:Nsinglets}
\end{table}
%%%%%%%%%%%%%%%%%%%%%%%%%%%%%%%%%%%%%%%%%%%%%%%%%%%%%%%%%%%%%%%%%%%
\end{widetext}

%%%%%%%%%%%%%%%%%%%%%%%%%%%%%%%%%%%%%%%%%%%%%%%%%%%%%%%%%%%%%%%%%%%%%%%%%%%%%%%%%%
%%%%%%%%%%%%%%%%%%%%%%%%%%%%%%%%%%%%%%%%%%%%%%%%%%%%%%%%%%%%%%%%%%%%%%%%%%%%%%%%%%

\end{document}